\documentclass[conference]{IEEEtran}
\IEEEoverridecommandlockouts
\usepackage{cite}
\usepackage{amsmath,amssymb,amsfonts}
\usepackage{algorithmic}
\usepackage{graphicx}
\usepackage{textcomp}
\usepackage{xcolor}
\usepackage{array}
\def\BibTeX{{\rm B\kern-.05em{\sc i\kern-.025em b}\kern-.08em
    T\kern-.1667em\lower.7ex\hbox{E}\kern-.125emX}}
\begin{document}

\title{Toward Mixed Reality Hybrid Objects \\ with IoT Avatar Agents
\thanks{Tri-Council of Canada, Canada Research Chairs Program.}
}

\author{\IEEEauthorblockN{Alexis Morris, Jie Guan, Nadine Lessio, Yiyi Shao}
\IEEEauthorblockA{\textit{Adaptive Context Environments Lab} \\
\textit{OCAD University}\\
Toronto, Canada \\
\{amorris, jguan, nlessio, yshao\}@faculty.ocadu.ca}
}
\begin{titlepage}

     \vspace{1cm}
        Full Citation: A. Morris, J. Guan, N. Lessio and Y. Shao, "Toward Mixed Reality Hybrid Objects with IoT Avatar Agents," 2020 IEEE International Conference on Systems, Man, and Cybernetics (SMC), Toronto, ON, Canada, 2020, pp. 766-773, doi: 10.1109/SMC42975.2020.9282914.

       \vspace*{1cm}

       \copyright2020 IEEE. Personal use of this material is permitted.  Permission from IEEE must be obtained for all other uses, in any current or future media, including reprinting/republishing this material for advertising or promotional purposes, creating new collective works, for resale or redistribution to servers or lists, or reuse of any copyrighted component of this work in other works.

       \vspace{1.5cm}
  
\end{titlepage}
\maketitle

\begin{abstract}
The internet-of-things (IoT) refers to the growing field of interconnected pervasive computing devices and the networking that supports smart, embedded applications. The IoT has multiple human-computer interaction challenges due to its many formats and interlinked components, and central to these is the need to provide sensory information and situational context pertaining to users in a more human-friendly, easily understandable format. This work addresses this by applying mixed reality toward expressing the underlying behaviors and states internal to IoT devices and IoT-enabled objects. It extends the authors’ previous research on IoT Avatars (mixed reality character representations of physical IoT devices), presenting a new head-mounted display framework and interconnection architecture. This contributes i) an exploration of mixed reality for smart spaces, ii) an approach toward expressive avatar behaviors using fuzzy inference, and iii) an early functional prototype of a hybrid physical and mixed reality IoT-enabled object. This approach is a step toward new information presentation, interaction, and engagement capabilities for smart devices and environments.
\end{abstract}


\section{Introduction}
\label{intro}

The Internet of Things (IoT) is generally used to describe technologies for networked embedded devices, which currently occupy many different environments, from adaptive smart homes to large-scale networked infrastructures used to monitor organizational sites, and even extends to governmental or city-scale applications\cite{gubbi2013internet}. To date much research has been centered on the technical applications of the IoT in industrial domains, but as the focus of these devices moves further into personal spaces, such as the home, designers and developers of IoT applications are faced with new human-computer interaction (HCI) challenges around 
user experience, and contextual awareness, particularly in terms of presenting the right information in the right format, and also in terms of increasing the methods of interaction with IoT systems \cite{nuamah2017human}, which have not seen as much attention in the academic and research communities. Concepts like context-awareness and contextually relevant presentation of information have now become essential parts in the everyday consumer use case of conventional IoT devices. 
 
Likewise, companies like Google and Amazon, for example, have taken strides to position smart-speakers with audio-based personal assistants like Google Assistant \footnote{Google Assistant: https://assistant.google.com/} and Alexa \footnote{Alexa: https://www.amazon.com/b?ie=UTF8\&node=17934671011}, as ``screenless'' operating systems for the management of IoT devices, wherein speech input is the primary interaction method. However, as homes, and lives become more connected, it can be expected that there will be a pressing need for more engaging interfaces and more information rich interaction modalities for the IoT that can better accommodate the complexities within a user's contextual lifecycle\cite{perera2013context}. 
While audio and speech affords much, as the scale of the embedded environment increases it can be expected that the use of speech-commands will easily become unwieldy. Likewise, common visual displays for IoT devices suffer from limitations of hardware, such as the impracticality of placing physical displays on each IoT object and physical input/output devices. Retro-fitting existing environment objects also represents further challenge. It is likely that these challenges will be addressed by a combination of technologies that incorporate audio, visual, and tangible interaction. 

In this case, Mixed Reality (MR) is poised to be an interesting gateway to more engaged interaction, as MR interfaces can blend both digital and real world environments into hybrid virtual and physical objects through recognition of objects, communication with object data states, and translation of those states into appropriate visual cues and representations for users. Mixed Reality, formally refers to ``a particular subset of Virtual Reality (VR) related technologies that involve the merging of real and virtual worlds somewhere along the virtuality continuum which connects completely real environments to completely virtual ones'' \cite{milgram1994taxonomy}. This includes the domain of Augmented Reality (AR), wherein ``the display of an otherwise real environment is augmented by means of virtual (computer graphic) objects'' \cite{milgram1994taxonomy} and its counterpart, Augmented Virtuality (AV), where a virtual environment is augmented with aspects of the real environment. Until recently, hardware limitations such as wearability, size, form-factor, and display field-of-view have prevented the rise of MR displays for ubiquitous computing, however, these are now becoming attainable for consumers \cite{lee2018interaction}. Currently AR has grown to include investment from some of the largest technical industry players such as Microsoft with their innovative Hololens series, and Apple with their series of AR enabled phones and a full fledged development kit \cite{norouzi2019systematic}. Funded startups like Magic Leap and other ventures into new heads up displays also point towards a larger future where AR becomes more adopted and engaged within people's daily lives \cite{norouzi2019systematic}. 

As the MR paradigm approaches consumer adoption in step with the IoT, opportunities arise to combine these technologies, leveraging AR smart glasses and common networking technologies, such as beacons or near-field communication. In terms of the interface issues mentioned above, and in the authors' previous research \cite{shao2019iot}, current IoT interfaces have also centered on dashboard approaches that are 2D and grid-themed, which allow for hierarchical arrangements, but still suffers in terms of providing rich data visualization, engagement, and interaction tools. This presents a need for MR and IoT interfaces to become more contextually linked and presented with more engaging, yet contextually relevant MR representations and form-factors. This is still an open area of research, as it combines IoT data with IoT overlayed 3D content models, characters, here referred to as IoT-Avatars, as in \cite{shao2019iot}, which serve to drive new interactions together with the physical IoT device or IoT-enabled objects. This work aims toward further exploration of the design and development of these hybrid objects and their interfaces. 

The objective is to humanize interaction with technical interfaces of conventional IoT devices, and to consider the shifting role of humans in the loop. This includes how to visualize IoT systems, and how those systems engage with humans, internal system components, and with other external systems. Working towards understanding these key relationships within an ever-increasing connected environment is an essential and complex interface challenge which can benefit from improved mechanisms (and better affordances) when engaged within a networked environment.
This work explores how an avatar for an IoT-enabled object (in this case a plant), can be implemented into a MR system, and how that avatar can interact from the technical system perspective while also providing a more expressive, engaging, and human-friendly approach to an IoT user interface. This work introduces the following key contributions: 
i) An exploration of a mixed reality design for smart objects,
ii) An approach toward expressive avatar behaviors using fuzzy inference,
iii) An early functional prototype of a hybrid physical and an MR IoT enabled object.

This section has provided a motivation to the IoT interaction challenges and potential for mixed reality IoT interfaces. Section \ref{background} highlights a background perspective on mixed reality in IoT research and mixed reality agent based research. Section \ref{core} proposes an extended architecture for IoT avatar characters based on the authors prior research. Section \ref{design} presents a design for an IoT avatar character AI system including fuzzy inference and visualization details. Section \ref{results} overviews the use case for this system and the system behavior results for an example scenario.
Section \ref{discussion} presents a brief discussion of the work. 
Section \ref{conclusion} summarizes the paper.

\section{Background: Mixed Reality Agents and the Internet-of-Things} 
\label{background}

The IoT is poised to play an important role in this new AR/MR enabled world by making engaging smart spaces a reality \cite{norouzi2019systematic} 
through devices like mobile phones, internet enabled products in homes, gaming systems that interact with haptics, and other smart devices. Together this blending of physical smart objects and aspects of AR form a ``Mirror World'', in which digital, physical, and social layers are strongly intertwined \cite{ricci2015mirror}. Smart spaces can exist on many different scales, from smaller spaces (e.g., living rooms), to entire cities \cite{ricci2015mirror}. One of the challenges with these new environments is defining a conceptual foundation, effective enough to model interactions with humans \cite{ricci2015mirror}. 
Currently handheld displays (smartphones) dominate the AR space, but there are new inroads being made into devices such as smart glasses, containing integrated music, smart agents like Alexa, gesture control, and notifications. Phones have also provided a platform for AR gaming with breakout games such as Pokemon GO \footnote{Pokemon Go: https://www.pokemongo.com/}, or Euclidean Skies (a platformer that floats in the air)\footnote{Euclidean Skies: http://euclideanskies.com/}, or more subtle applications like WallaMe which allow users to leave virtual messages for one another at real locations in the world \footnote{WallaMe: http://walla.me/}. Moving more into an MR space, companies like Moment Factory 
have been drawing from both MR and interactive theatre worlds, to create shared emotional, immersive experiences in public space \cite{fouquet2017digital} through the use of smart objects, large scale projection, and site specific installations to tell highly engaging interactive stories.


In terms of agents, a mixed reality agent, otherwise known as an MiRA, can be defined as hardware or software based entities that have the conventional agent systems theory attributes: Autonomy, Social Ability, Reactivity, and Pro-Activity \cite{wooldridge1995intelligent} and that exist with a mixture of virtual and physical embodiment (usually with stronger virtual embodiment versus physical embodiment) \cite{holz2011mira}. In particular, the dimensions of \emph{agency} can be combined with dimensions of \emph{interactive capacity}, and \emph{corporeal presence}, as defined in \cite{holz2011mira}, providing a method to consider a spectrum of MiRA implementations. In this, embodiment can be considered in different ways; one way MiRAs can do this is to demonstrate structural coupling, in which a system is embodied if it is able to sense, affect and be affected by its environment \cite{quick1999bots, holz2011mira}. They can also have social embodiment, in which an agent can interact with users and other agents within a social framework or environment \cite{holz2011mira, dourish2004action}. This highlights that intelligent agents must engage with its environment in some fashion, whether through sensors, actuators, response to users, or changes in behaviour based on environmental or human factors.

Consumer products like Alexa, can be considered within the MiRA dimensions, as they do have some level of social ability (voice interaction), and can control physical objects. However, even more embedded projects, like Kobito Virtual Brownies \cite{aoki2005kobito}, present stronger MiRA examples. In the latter case, imaginary creatures interact with the real world by moving real objects through a system composed of an overhead camera utilizing computer vision, and a table with a magnetic caddy under it for manipulating with real world physical objects on the surface. As a result the virtual agents move real objects while users interact with the virtual agents through a tablet based viewport (without using markers, head- mounted displays, or wearable sensors). In contrast, purely physically embodied agents like Paro \cite{parochang2013situated}, a therapy robot developed by the Japanese company AIST, have also been designed with an embodied physical presence, in Paro's case as a baby seal. Here, the physicality of robots encourages anthropomorphism, and the exchange of emotional responses between agents and users, and indicates potential similar effects for embodiment within virtual agent systems, particularly where they can be merged with the IoT.

\begin{figure*}[h]
	\centering
		\includegraphics[width=5.25in]{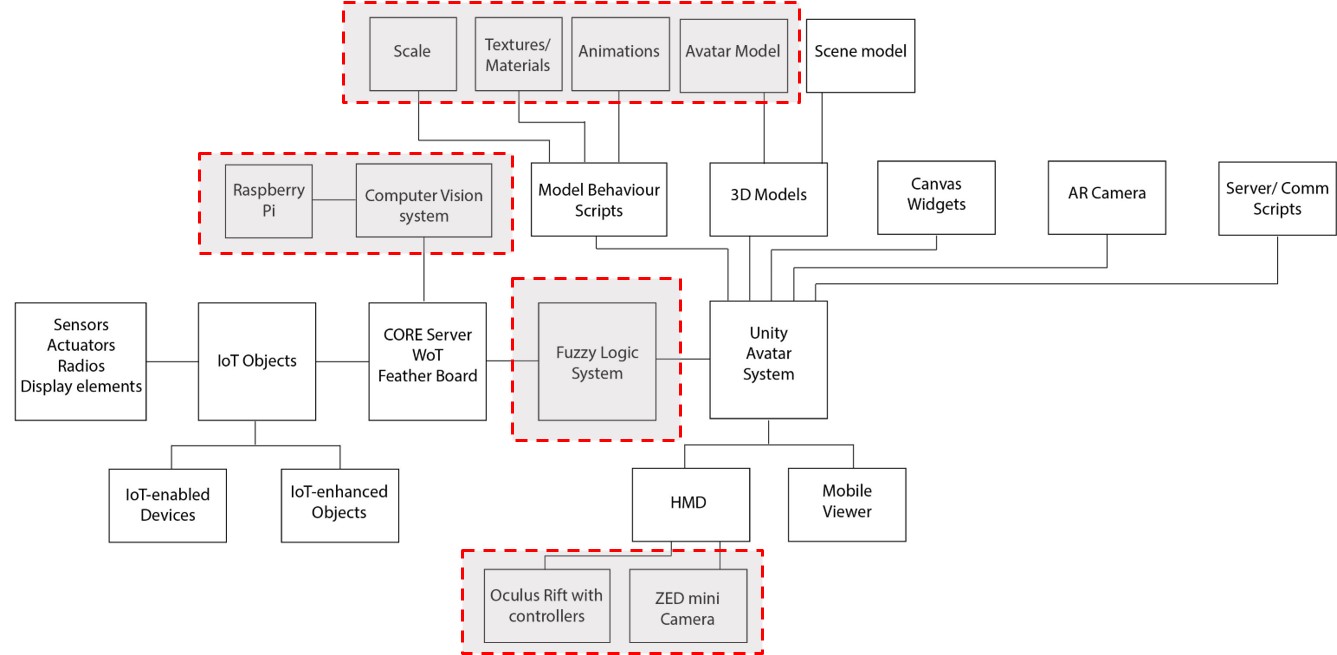}
	\caption{Extending the Contextual Reality (CoRe) framework for IoT Avatars, based on \cite{shao2019iot}. Extensions to CoRe include a fuzzy inference system, computer vision system, new avatar models, script events, and head-mounted mixed reality components.}
	\label{fig:extended framework}
\end{figure*}

\section{Extending a CoRe Architecture for Mixed Reality IoT Avatars}
\label{core}

\begin{figure}[h]
	\centering
		\includegraphics[width=\linewidth]{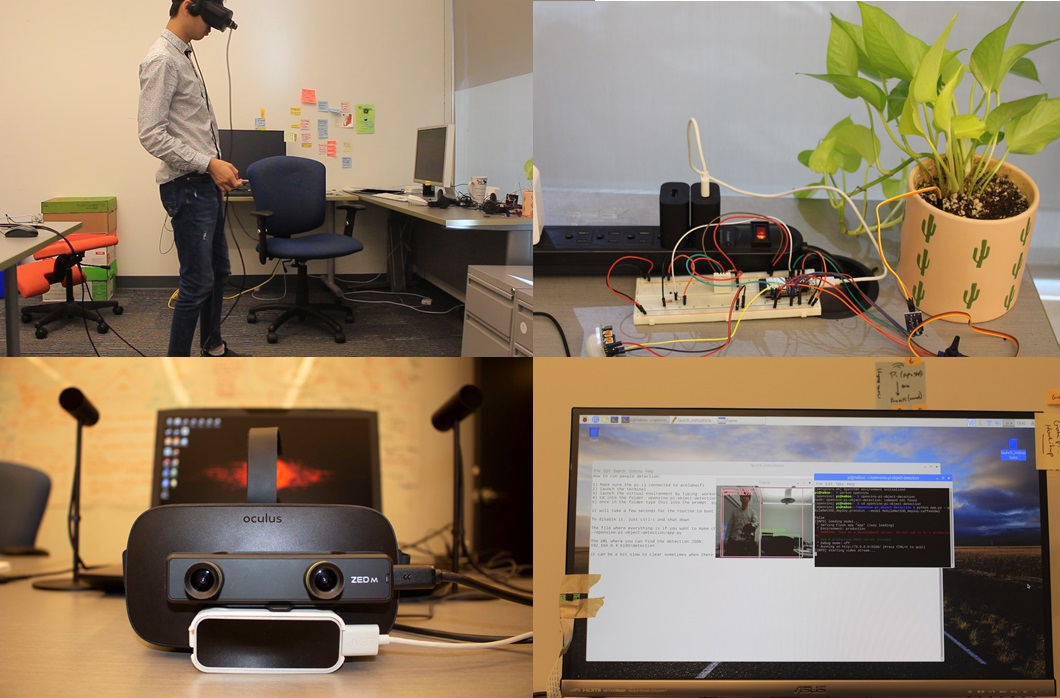}
	\caption{Hardware components for a mixed reality HMD, including Oculus and Zed Mini cameras, in connection with IoT components (Arduino, Raspberry Pi).}
	\label{fig:Hardware componnets}
\end{figure}

In the authors' previous research \cite{shao2019iot} an IoT Avatar system was developed using a mobile AR framework, for a simple hand-held MR early concept scenario, and is derived from the authors' Contextual Reality (CoRe) architecture of \cite{morris2018}. This work added preliminary research into exploring mixed reality visualization in smart spaces. However, the approach was focused solely on mobile phone AR technologies and the presentation of designs for the IoT Avatar concept, for a smart plant scenario. It hinted at the potential to extend toward visualization with immersive head-mounted displays (HMDs), refinements of the avatar system, and potential user studies engaging and encouraging human interactions with the goal of improving the relationship between users and avatars. In this work, the CoRe architecture has been extended, in Figure \ref{fig:extended framework}, to include: i) a Mixed Reality head-mounted display approach, ii) better sensor interaction, iii) more adaptive behaviors of the avatar, iv) more engaging presentations of the avatar, v) and more expressive avatar features and feedback through emotional representation; each as described below. Figure \ref{fig:Hardware componnets} presents these hardware components for the extended IoT Avatar System.

\textbf{Head-mounted Display (HMD):} 
In selecting an immersive HMD approach, a video-see through approach was considered. 
This combined a ZED Mini camera and an Oculus Rift VR headset, with Oculus Touch controllers as the Mixed Reality hardware. The ZED Mini Camera, attached in front of the Oculus Rift headset, captures real-time video (with depth data) for displaying virtual content \footnote{https://www.stereolabs.com/zed-mini/setup/rift/}. While this approach has the limitations of being tethered to a PC, having to use hand controllers, and a low-resolution display, it allows users of the HMD to experience a relatively immersive presentation of visualized content in-situ (with stereoscopic depth, environment tracking, head-tracking, positioning, and scale) and anchored to the physical environment.  

\textbf{Data Sensors:}
For the selected scenario, Arduino sensors were incorporated, namely a Photoresistor GL5528 to detect the brightness value of the space, a Capacitive Soil Moisture Sensor to capture the soil moisture level of a plant, and a Raspberry Pi running OpenCV, a camera, and a machine learning module to detect how many people are visible in the environment. An Arduino ESP32 hosts this sensor data, while a Raspberry Pi hosts the camera data, making these available to the network through request-response frameworks. 

\textbf{Unity Engine:}
Unity3D provides the visualization engine for presenting the avatar, its behaviours, and also logic for the emotion of the avatar. It requests the emotional results from a fuzzy inference system, running on a local server in the environment, and based on its state the avatar will be visualized, having diverse animations, materials, particle effects, and scales. 
For example, when the avatar is in a Happy state, it can dance, change to a smile texture, generate colorful and dynamic particles, or change its size.

\textbf{Fuzzy Inference System:}
The values from the sensor system are inputs for calculating emotional states, which are obtained using a measure of arousal and valence scores attained through use of a fuzzy inference system. The environment can be modified by turning on and off its lights, opening or closing a curtain, adding water to the plant or allowing it to dry, and by people moving toward or out of the visible range near the plant.

\section{IoT Avatar: Agent Design}
\label{design}

\begin{figure}
	\centering
		\includegraphics[width=\linewidth]{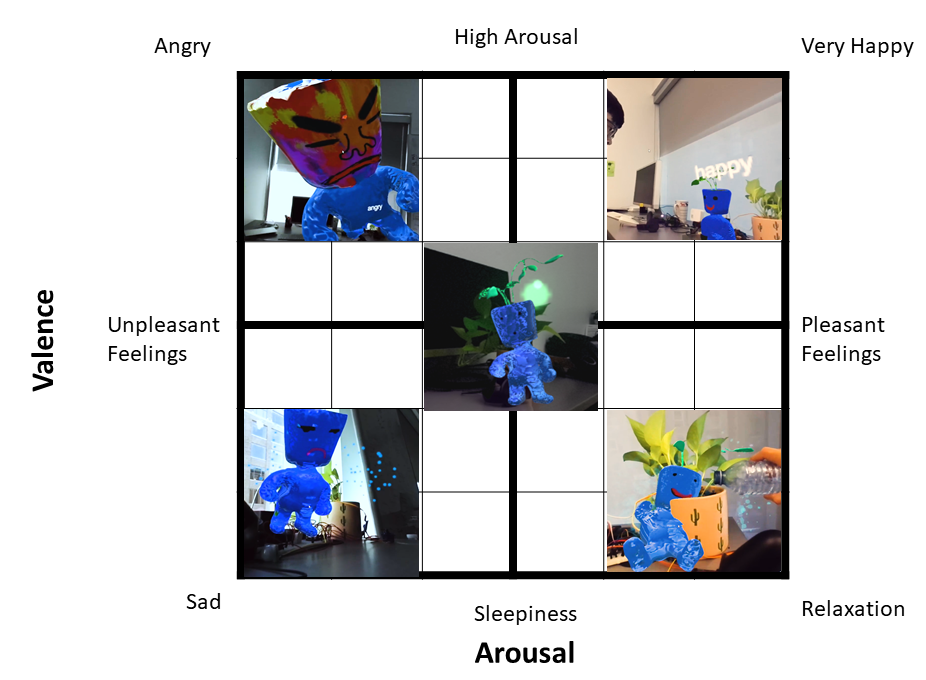}
	\caption{Mapping process for applying arousal and valence into useful emotional states, as in \cite{guan2020exploring}, based on \cite{mandryk2007fuzzy}. 
	}
	\label{fig:Mapping}
\end{figure}

\begin{figure}[h]
	\centering
		\includegraphics[width=\linewidth]{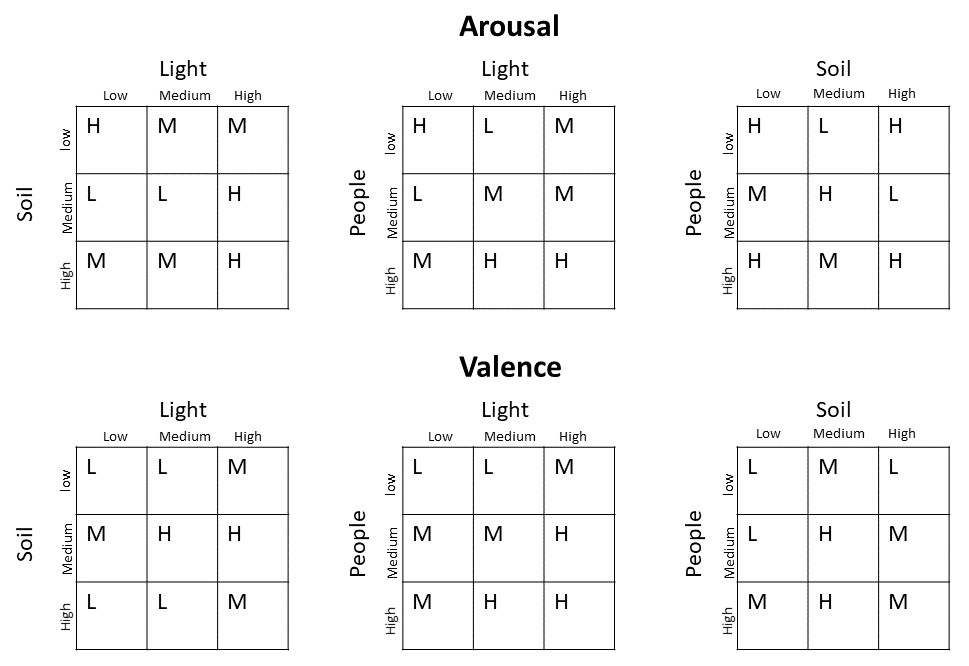}
	\caption{Relationships between arousal and valence, based on levels of soil, light, and moisture readings (Low, Medium, or High). These are subjectively provided, and can be further adapted in future studies.}
	\label{fig:Relationship}
\end{figure}

The extended components above have been used to design a functional MR IoT Avatar system; an embodied virtual agent with engaging behavior, which communicates to users through emotional states, as a combination of arousal and valence, which are calculated via a fuzzy inference system, as in Figures \ref{fig:Mapping} and \ref{fig:Relationship}. 

\subsection{An IoT Avatar for a Smart-Plant Hybrid Object}

\begin{figure}[h]
	\centering
		\includegraphics[width=\linewidth]{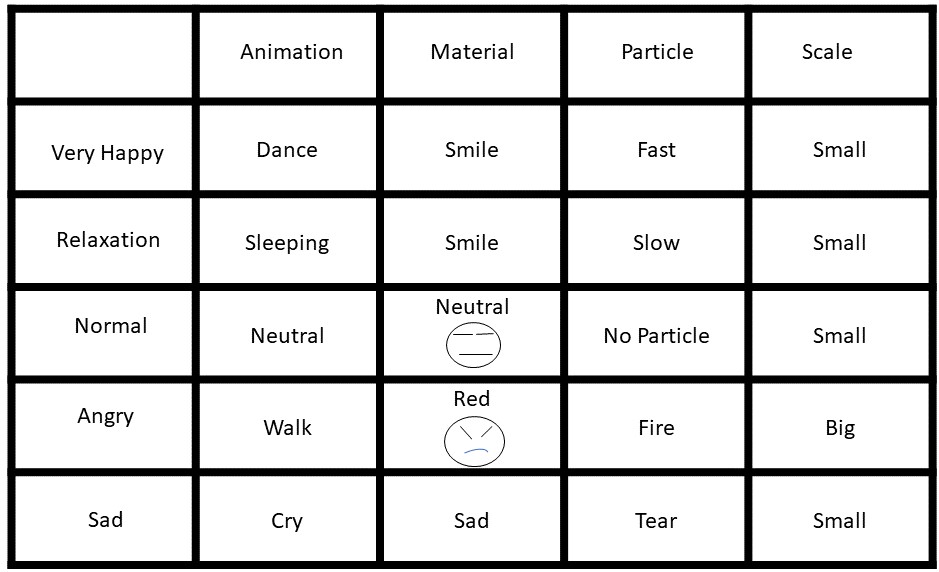}
	\caption{Properties of engagement behaviors, within Unity3D, placeholders for emotional activity.}
	\label{fig:properties}
\end{figure}

In this work, a Plant avatar is the focus, considering the underlying question of whether an MR IoT Avatar can help to make interactions with a house plant more engaging and immersive, toward improving the relationship between humans in the loop, the physical plant object, and the virtual avatar. As such, the presentation of the avatar character have been designed to fit this theme, with states as shown in Figure \ref{fig:properties}. 

In terms of design consideration, the avatar is designed solely to express multiple arousal-valence emotional states representing the behavior of a plant with animation behaviors, textures and materials for facial expression, particle effects, and state changes, especially, changes in scale (such as for an angry state). The avatar should be anchored to the physical object as closely as possible, and also represent the object in terms of shape and structure, while also having a default size similar to the object. Further, the avatar should be reactive to the environment, such as changes to the lighting of the background scene should affect the conditions of the virtual avatar. Lastly, for conventional interaction and communication purposes for this iteration, the avatar can only communicate using text descriptors and may also provide interactive buttons and other widgets.

When the avatar in Happy state, it can dance in the space with smiling textures and colorful, fast particles. This action aims to engage and reward users with an exciting response from the avatar's environment. When the avatar is in the Relaxation state, it displays a going-to-sleep animation and displays water-like particles in blue color, allowing users to be aware of the avatar's comfortable experience. When the avatar is in an Angry state, likewise, the avatar is designed to grow large in scale, and to fill up the room with red textures and fire particles, as it attempts to obtain attention from the users (on behalf of the plant). Lastly, when in a Saddened state, the avatar can express this by a shaking head animation, frown texture, and tear particles, encouraging users to address this state by tending to the plant's needs.

\subsection{Fuzzy Inference System}

\begin{figure}[h]
	\centering
		\includegraphics[width=\linewidth]{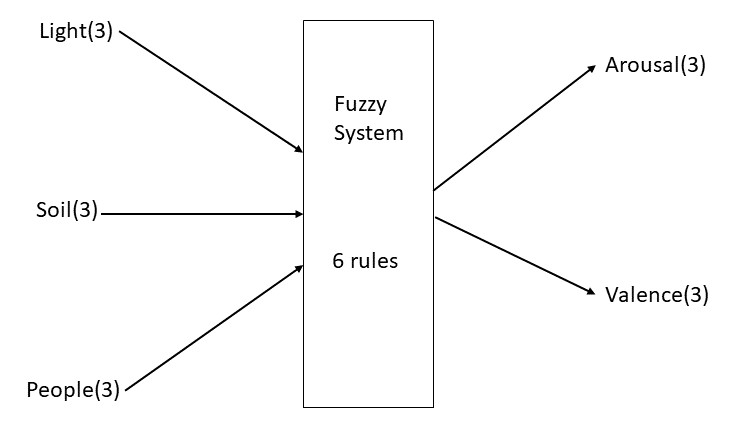}
	\caption{Input signals for a fuzzy inference system calculating Arousal and Valence, based on Soil, Light, and Moisture levels, based on \cite{mandryk2007fuzzy}.}
	\label{fig:Input signals}
\end{figure}

\begin{figure}[h]
	\centering
		\includegraphics[width=\linewidth]{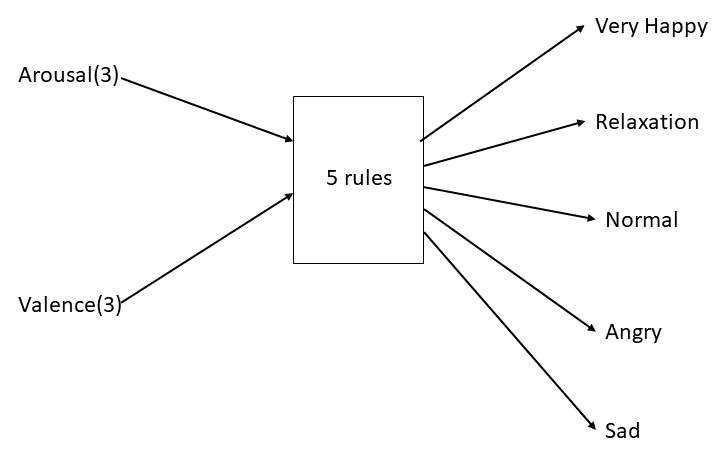}
	\caption{Emotion state related to arousal and valence}
	\label{fig:Emotion State}
\end{figure}

The fuzzy inference system provides a convenient way to convert sensor inputs, which have a wide range of values, into ``reasonable'' cross-sections of responses based on ranges of each input. Inputs in this fuzzy system are based on the information from the environment, including brightness, soil moisture, and the number of people present. The value of these data convert into Poor, Average, and Good via the auto-membership functions provided by Skfuzzy\footnote{https://pythonhosted.org/scikit-fuzzy/}. The consequent (outputs) from this fuzzy logic system, including Arousal and Valence, range between Low, Medium, and High. 
There are six compound rules for the combination of Arousal and Valence, which contribute five emotion state outputs, as in Figure \ref{fig:properties}. These rules are created based on a combination of light and soil, light and people, and soil and people. In each case, arousal and valence values are considered as either normal, low, or high, and the following states are provided for visualization in Unity: 

1.valence is High and arousal is Low: emotion = 'relaxation'

2.valence is High and arousal is High: emotion = 'happy'

3.valence is Low and arousal is High: emotion = 'angry'

4.valence is Low and arousal is Low: emotion = 'sad'

5.else: emotion = 'normal'

These have been combined, as seen in Figure \ref{fig:Relationship}, based on subjective estimates of how each combination would effect the state of a plant, and how that could be interpreted as low, medium, or high arousal or valence. These rules are presented below (note that a more realistic or optimized behavior controller for the system is an area for future study): 

In terms of Rule 1, the levels of soil (low, med, high) to levels of light (low, med, high) are subjectively compared in order to produce a score of Arousal (low, med, high). This is seen in the top-left group of Figure \ref{fig:Relationship}.
In terms of Rule 2, the number of people (low, med, high) to levels of light (low, med, high) are subjectively compared in order to produce a score of Arousal (low, med, high). This is seen in the top-middle group of Figure \ref{fig:Relationship}.
In terms of Rule 3, the number of people (low, med, high) to levels of soil (low, med, high) are subjectively compared in order to produce a score of Arousal (low, med, high). This is seen in the top-right group of Figure \ref{fig:Relationship}. 
In terms of Rule 4, the levels of soil (low, med, high) to levels of light (low, med, high) are subjectively compared in order to produce a score of Valence (low, med, high). This is seen in the down-left group of Figure \ref{fig:Relationship}. 
In terms of Rule 5, the number of people (low, med, high) to levels of light (low, med, high) are subjectively compared in order to produce a score of Valence (low, med, high). This is seen in the down-middle group of Figure \ref{fig:Relationship}. 
In terms of Rule 6, the number of people (low, med, high) to levels of soil (low, med, high) are subjectively compared in order to produce a score of Valence (low, med, high). This is seen in the down-right group of Figure \ref{fig:Relationship}. 
Together these map well into the behavioral states of the avatar system, as in Figure \ref{fig:Mapping} and \ref{fig:Input signals}.
The reasonable max and min of these values are gathered and converted into a percentage. The range of \emph{brightness} is 0 to 780, soil \emph{moisture} is from 1800 to 3100, and the maximum number of \emph{people} in the scenario is 4. Combining these into the fuzzy system produces \emph{arousal} and \emph{valence}, with a series of zones used to present the emotional state for the avatar.


\section{Results} 
\label{results}

\begin{figure}[h]
	\centering
		\includegraphics[width=\linewidth]{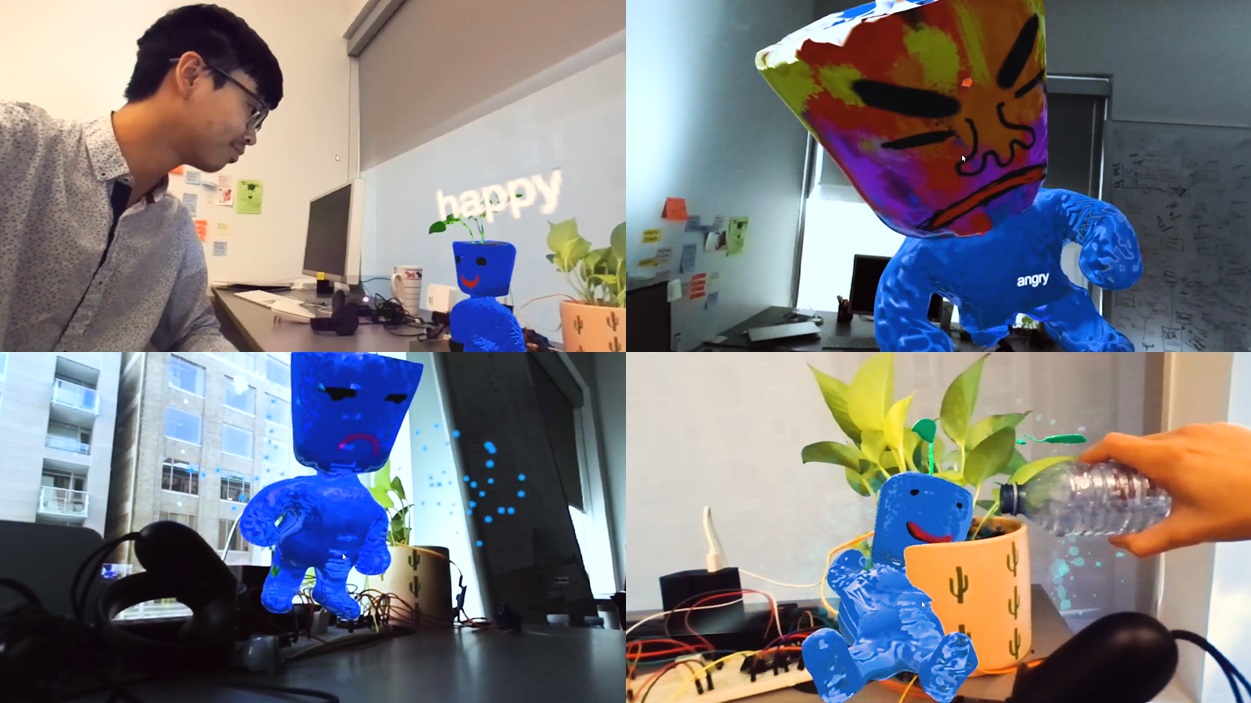}
	\caption{Information was gathered for a simple plant IoT-Avatar scenario, modifying the avatar behaviors through changes in the environment states (lighting, moisture, presence of people).} 
	\label{fig:Information}
\end{figure}

Figure \ref{fig:Information} shows the resulting IoT avatar design in context as a representation of how the approach leads to a more immersive and engaging information representation of a hybrid object, in this case a smart-plant (video of this IoT plant avatar is available online \cite{guan2020exploring} \footnote{https://www.youtube.com/watch?v=o80bpAzf24E}). This outcome is considered more immersive than a conventional 2D display representation, and even moreso than a mobile AR display, as in \cite{shao2019iot}, although this remains to be tested in-depth.

\begin{figure*}[h]
	\centering
		\includegraphics[width=6.5in]{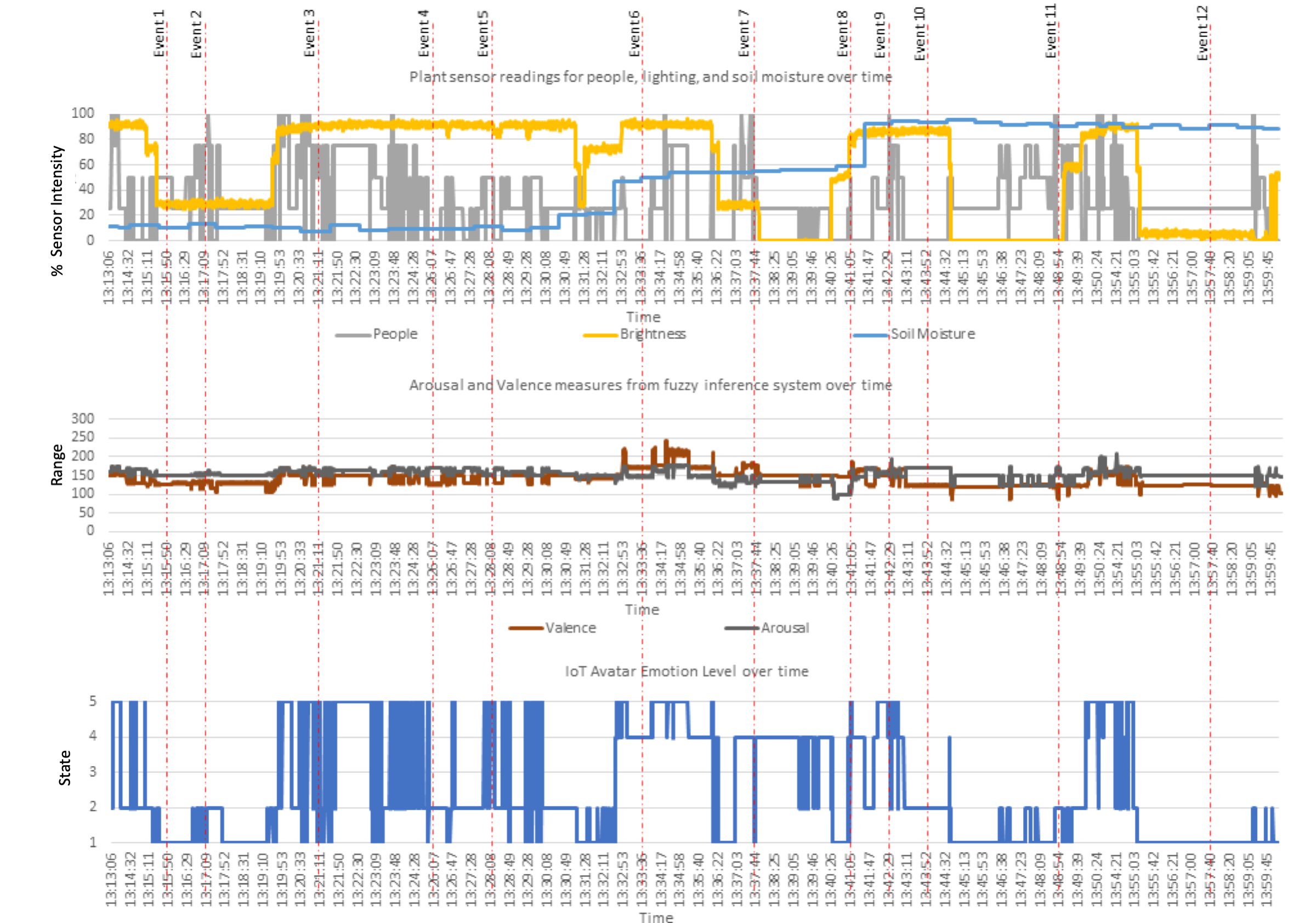}
	\caption{Sensor events from the physical plant object (i.e., manipulating lighting, soil moisture, or people present) are communicated to the mixed reality plant avatar and expressed as one of its five emotional states. Twelve events modifying these variables are shown across a fixed time period in order to investigate functional responsiveness of the avatar (via its arousal-valence emotion states). }
	\label{fig:Investigating}
\end{figure*} 

The current section presents a small-scale functional evaluation of system behaviors, in which twelve representative events are selected to consider the functionality of the system when events take place in the IoT system (see Figure \ref{fig:Investigating}). These are shown in data gained from running the system across a single hour to demonstrate how the avatar state relates to the environment data. A more detailed exploration remains for future research, as the present focus is on whether the avatar system appropriately presents states based on the situation of the plant and the environmental context. 
In this, the ranges of Arousal and Valence are from 0 to 300 representing the output of the membership values in the fuzzy logic system. These values, once obtained, as in Figure \ref{fig:Arousal and Valance}, are mapped based on a simple test of whether it is above the average (hence, if the score is more than 150 then it is simply considered as high arousal, or low arousal if less than 150). This simple mapping converts arousal and valence for presenting cross sections identified above (and will be adjusted in future work toward a more realistic model). 

\begin{figure}[h]
	\centering
		\includegraphics[width=\linewidth]{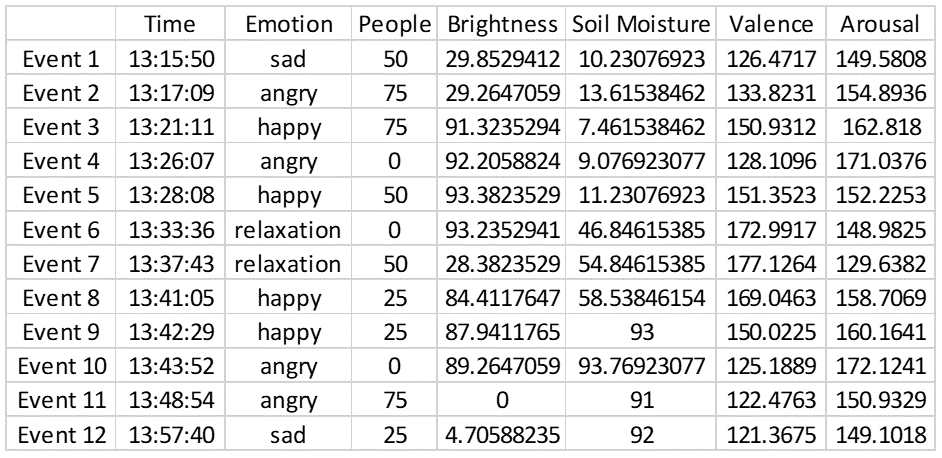}
	\caption{Data table showing specific events of interest used to show Avatar Behavior, over a fixed time period. The emotional state of the avatar is shown per event, as a mapping of arousal and valence scores, where arousal and valence are calculated based on the percentage of people present (range 0-3), and the sensor readings for room brightness and soil moisture.}
	\label{fig:Data table}
\end{figure}

\begin{figure}
	\centering
		\includegraphics[width=\linewidth]{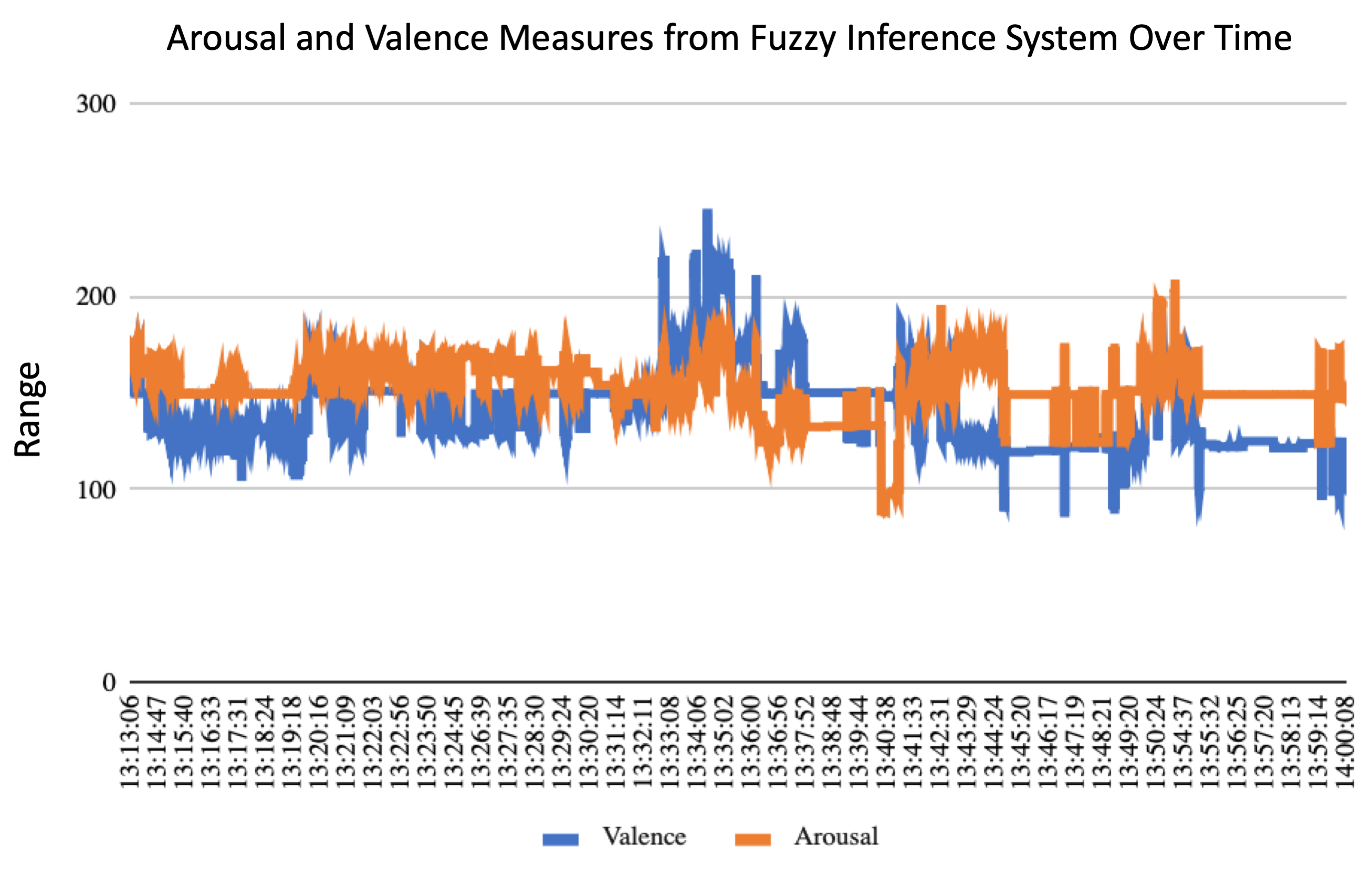}
	\caption{Arousal and Valance measures as outputs from the fuzzy inference system (based on soil, light, and the presence of people) over a fixed time period, during specific events (as in Figures \ref{fig:Investigating} and \ref{fig:Data table}).}
	\label{fig:Arousal and Valance}
\end{figure}

The emotional states shown (Figure \ref{fig:Investigating}) are as follows: state 1 represents sad, state 2 is angry, state 3 is normal, state 4 is relaxation, and state 5 is happy. A selection of events over the time-period is shown in Figure \ref{fig:Data table}, allowing for considering the avatar behavior versus the environment changes. For Example, in Event 1, there are two people in the room, so this value state is in the medium; the brightness is low since only one light is turned on in the room; also, in this case the plant was not watered for three days, allowing the soil moisture to be in the low state. As a result, the fuzzy logic system determines that the valence and arousal would be both in the low state, and this leads to the sad emotional state. Likewise, in Event 6, by adding water to the plant, so soil moisture is medium, followed by increasing the light level, with no people in vicinity of the plant, 
results in the avatar displaying the relaxation state. Also, in Event 11, by adding too much water to the plant, and completely turning off all the light while also having four people present near the plant, 
generated low valence and high arousal, and this leads to the angry emotional state. 

Together, these events show a reasonable functional behavior model of the IoT Avatar and its reactiveness to environmental states, which are aimed at evoking user awareness with behaviors that could lead to more engagement with environmental objects. However, as indicated, a detailed evaluation of the interaction and effectiveness of this approach is left for future stages of this research.



\section{Discussion}
\label{discussion}


Despite growth into everyday computing areas, the IoT is still faced with HCI challenges around user experience, and contextual awareness. This includes how to visualize the systems and subsystems in the IoT, how to dictate commands to those systems, and how to engage with these systems and with other users, from a multi personal and multi-agent system perspective; whether from human-to-human or human-to-technology, or the interactions between technological components. The avatar system presented has provided the groundwork for exploring these relationships with hybrid mixed reality and IoT-enabled objects. Table \ref{tab:1} shows how the proposed approach compares to several related work, with a more thorough analysis left for future stages of the work (for instance, \cite{holz2011mira} shows multiple approaches to be compared). In this the CoRe design considerations of \cite{morris2018} for a multidimensional IoT framework can be used as an early comparison criteria, indicating that even more needs to be done to create IoT devices that have virtual, ambient, collaborative, contextual, inferential, and networking capabilities. This work looks toward this form of IoT environment design gaining more maturity in the future. 

\begin{table*}[!htb]
  \centering
\begin{tabular}{ | m{6em} | m{1.3cm}| m{2cm} | m{1.5cm} | m{2cm} |m{2cm} |m{2cm} |}
\hline
\textbf{CoRe Design Considerations \cite{morris2018}} & \textbf{Virtual} &\textbf{Ambient} & \textbf{Collaborative} & \textbf{Context-Aware} &\textbf{Inferential} & \textbf{Networking} \\ 
\hline
IoT Avatar 2.0 (Current Prototype) & Yes, HMD with ZED Camera attached. &  Yes, individuals can manipulate the information of the context to effect the emotion of the plant avatar. & No. & Yes, the system contains soil moisture sensor, light sensor, and computer vision model to detect the number of people in the environment.  & Yes, fuzzy logic can be extended into machine learning system. & Yes, using Flask-SocketIO and HTTP request for data communication. \\ 
\hline
IoT Avatar 1.0 (Early Prototype) \cite{shao2019iot} & Yes, mobile AR. & Yes, users are able to press buttons on the mobile screen to toggle the LED light and Servo Motor  & No. & No. & No. & Yes, using server based forms for JSON data. transmission. \\ 
\hline
Ambient Bot\cite{gushima2017ambient} & Yes, HMD and Webcam used. & Yes, the creature only speaks and displays information while the user focuses on it. & No. & No. & No. & No. \\ 
\hline
Welbo\cite{anabuki2000welbo} & Yes, HMD with sensor attached. &  No.  & No. & Yes, the system includes speech recognition module and perceives the user's movement.  & No. & No. \\ 
\hline
Kobito\cite{aoki2005kobito} & Yes, tablet with physical interaction. & Yes, people interact with Kobito through moving the real objects. & No. & No. & No. & No. \\ 
\hline


\end{tabular}
\caption{Comparing representative IoT Avatar systems to the CoRe \cite{morris2018} design considerations.}
  \label{tab:1}
\end{table*}

\section{Conclusion}
\label{conclusion}

The domain of mixed reality is beginning to merge with designs for consumer-driven smart environments enabled by the internet of things. This provides significant user interaction challenges and opportunities to develop engaging and expressive smart environments that can broaden and improve the interfaces of IoT applications, toward increasing their effectiveness. In this work a functional prototype of a hybrid mixed reality IoT-enabled object, and an architecture involving a fuzzy inference system agent has been shown, as a step toward creating these future interface mechanisms. While this work does not evaluate the effectiveness of such approaches to information visualization, this is a direction for future research, and the mixed reality and smart environment communities are encouraged to pursue the concept further, toward understanding the benefits of embodied virtual avatar agents as multi-modal interfaces in practical smart environments.

\section*{Acknowledgment}
This work gratefully acknowledges funding from the Tri-council of Canada under the Canada Research Chairs program.

\bibliographystyle{IEEEtran}
\bibliography{references}

\begin{thebibliography}{10}
\providecommand{\url}[1]{#1}
\csname url@samestyle\endcsname
\providecommand{\newblock}{\relax}
\providecommand{\bibinfo}[2]{#2}
\providecommand{\BIBentrySTDinterwordspacing}{\spaceskip=0pt\relax}
\providecommand{\BIBentryALTinterwordstretchfactor}{4}
\providecommand{\BIBentryALTinterwordspacing}{\spaceskip=\fontdimen2\font plus
\BIBentryALTinterwordstretchfactor\fontdimen3\font minus
  \fontdimen4\font\relax}
\providecommand{\BIBforeignlanguage}[2]{{%
\expandafter\ifx\csname l@#1\endcsname\relax
\typeout{** WARNING: IEEEtran.bst: No hyphenation pattern has been}%
\typeout{** loaded for the language `#1'. Using the pattern for}%
\typeout{** the default language instead.}%
\else
\language=\csname l@#1\endcsname
\fi
#2}}
\providecommand{\BIBdecl}{\relax}
\BIBdecl

\bibitem{gubbi2013internet}
J.~Gubbi, R.~Buyya, S.~Marusic, and M.~Palaniswami, ``Internet of things (iot):
  A vision, architectural elements, and future directions,'' \emph{Future
  generation computer systems}, vol.~29, no.~7, pp. 1645--1660, 2013.

\bibitem{nuamah2017human}
J.~Nuamah and Y.~Seong, ``Human machine interface in the internet of things
  (iot),'' in \emph{2017 12th System of Systems Engineering Conference
  (SoSE)}.\hskip 1em plus 0.5em minus 0.4em\relax IEEE, 2017, pp. 1--6.

\bibitem{perera2013context}
C.~Perera, A.~Zaslavsky, P.~Christen, and D.~Georgakopoulos, ``Context aware
  computing for the internet of things: A survey,'' \emph{IEEE communications
  surveys \& tutorials}, vol.~16, no.~1, pp. 414--454, 2013.

\bibitem{milgram1994taxonomy}
P.~Milgram and F.~Kishino, ``A taxonomy of mixed reality visual displays,''
  \emph{IEICE TRANSACTIONS on Information and Systems}, vol.~77, no.~12, pp.
  1321--1329, 1994.

\bibitem{lee2018interaction}
L.-H. Lee and P.~Hui, ``Interaction methods for smart glasses: A survey,''
  \emph{IEEE Access}, vol.~6, pp. 28\,712--28\,732, 2018.

\bibitem{norouzi2019systematic}
N.~Norouzi, G.~Bruder, B.~Belna, S.~Mutter, D.~Turgut, and G.~Welch, ``A
  systematic review of the convergence of augmented reality, intelligent
  virtual agents, and the internet of things,'' in \emph{Artificial
  Intelligence in IoT}.\hskip 1em plus 0.5em minus 0.4em\relax Springer, 2019,
  pp. 1--24.

\bibitem{shao2019iot}
Y.~Shao, N.~Lessio, and A.~Morris, ``Iot avatars: Mixed reality hybrid objects
  for core ambient intelligent environments,'' \emph{Procedia Computer
  Science}, vol. 155, pp. 433--440, 2019.

\bibitem{ricci2015mirror}
A.~Ricci, M.~Piunti, L.~Tummolini, and C.~Castelfranchi, ``The mirror world:
  Preparing for mixed-reality living,'' \emph{IEEE Pervasive Computing},
  vol.~14, no.~2, pp. 60--63, 2015.

\bibitem{fouquet2017digital}
L.~Fouquet, ``A digital campfire in a multimedia hive: Moment factory,''
  \emph{ETC MEDIA}, no. 111, pp. 30--35, 2017.

\bibitem{wooldridge1995intelligent}
M.~Wooldridge and N.~R. Jennings, ``Intelligent agents: Theory and practice,''
  \emph{The knowledge engineering review}, vol.~10, no.~2, pp. 115--152, 1995.

\bibitem{holz2011mira}
T.~Holz, A.~G. Campbell, G.~M. O’Hare, J.~W. Stafford, A.~Martin, and
  M.~Dragone, ``Mira—mixed reality agents,'' \emph{International journal of
  human-computer studies}, vol.~69, no.~4, pp. 251--268, 2011.

\bibitem{quick1999bots}
T.~Quick, K.~Dautenhahn, C.~L. Nehaniv, and G.~Roberts, ``On bots and bacteria:
  Ontology independent embodiment,'' in \emph{European Conference on Artificial
  Life}.\hskip 1em plus 0.5em minus 0.4em\relax Springer, 1999, pp. 339--343.

\bibitem{dourish2004action}
P.~Dourish, \emph{Where the action is: the foundations of embodied
  interaction}.\hskip 1em plus 0.5em minus 0.4em\relax MIT press, 2004.

\bibitem{aoki2005kobito}
T.~Aoki, T.~Matsushita, Y.~Iio, H.~Mitake, T.~Toyama, S.~Hasegawa, R.~Ayukawa,
  H.~Ichikawa, M.~Sato, T.~Kuriyama \emph{et~al.}, ``Kobito: virtual
  brownies,'' in \emph{ACM SIGGRAPH 2005 emerging technologies}, 2005, pp.
  11--es.

\bibitem{parochang2013situated}
W.-L. Chang, S.~{\v{S}}abanovi{\'c}, and L.~Huber, ``Situated analysis of
  interactions between cognitively impaired older adults and the therapeutic
  robot paro,'' in \emph{International Conference on Social Robotics}.\hskip
  1em plus 0.5em minus 0.4em\relax Springer, 2013, pp. 371--380.

\bibitem{morris2018}
A.~Morris and N.~Lessio, ``Deriving privacy and security considerations for
  core: An indoor iot adaptive context environment,'' in \emph{Proceedings of
  the 2nd International Workshop on Multimedia Privacy and Security}.\hskip 1em
  plus 0.5em minus 0.4em\relax ACM, 2018, pp. 2--11.

\bibitem{guan2020exploring}
J.~Guan, N.~Lessio, Y.~Shao, and A.~Morris, ``Exploring a mixed reality
  framework for the internet-of-things: Toward visualization and interaction
  with hybrid objects and avatars,'' in \emph{2020 IEEE Conference on Virtual
  Reality and 3D User Interfaces Abstracts and Workshops (VRW)}, 2020, p. 858.

\bibitem{mandryk2007fuzzy}
R.~L. Mandryk and M.~S. Atkins, ``A fuzzy physiological approach for
  continuously modeling emotion during interaction with play technologies,''
  \emph{International journal of human-computer studies}, vol.~65, no.~4, pp.
  329--347, 2007.

\bibitem{gushima2017ambient}
K.~Gushima, H.~Akasaki, and T.~Nakajima, ``Ambient bot: delivering daily casual
  information through eye contact with an intimate virtual creature,'' in
  \emph{Proceedings of the 21st International Academic Mindtrek Conference},
  2017, pp. 231--234.

\bibitem{anabuki2000welbo}
M.~Anabuki, H.~Kakuta, H.~Yamamoto, and H.~Tamura, ``Welbo: An embodied
  conversational agent living in mixed reality space,'' in \emph{CHI'00
  extended abstracts on Human factors in computing systems}, 2000, pp. 10--11.

\end{thebibliography}

\end{document}